\documentclass{PoS}
\usepackage{amssymb,bbm,amstext,amsmath,amsthm,amsfonts,amsmath,amscd,relsize,url,mathtools,booktabs}
\newcommand{\p}{\partial}
\newcommand{\pa}{\partial}
\newcommand{\pt}{\tilde{\partial}}
\newcommand{\rd}{\mathrm{d}}
\newcommand{\dd}{\mathrm{d}}

\newcommand{\id}{\mathbbm{1}}

\newcommand{\n}{\nabla}					
\newcommand{\nc}{\nabla^c}			
\newcommand{\lc}{\mathring{\n}}		

\newcommand{\be}{\begin{equation}}
\newcommand{\ee}{\end{equation}}
\newcommand{\bea}{\begin{eqnarray}}
\newcommand{\eea}{\end{eqnarray}}

\newcommand{\Lt}{{\tilde{L}}}
\newcommand{\Pt}{{\tilde{P}}}
\newcommand{\Hh}{\mathcal{H}}

\newcommand{\HH}{\mathcal{H}}

\newcommand{\Th}{{\mathcal{T}}}
\newcommand{\PS}{\mathcal{P}}
\newcommand{\TT}{T\oplus T^*}

\newcommand{\Lie}{\mathcal L}

\newcommand{\ap}{\alpha}
\newcommand{\bt}{\beta}

\newcommand{\bracd}{\bl\ ,\ \br}
\newcommand{\se}{\Gamma}
\newcommand{\bl}{[\![}
\newcommand{\br}{]\!]}
\newcommand{\la}{\langle}
\newcommand{\ra}{\rangle}
\def\to{\rightarrow}
\def\tl{\tilde}

\newcommand{\xt}{\tl{x}}

\newcommand{\nB}{\n^{\mathrm{B}}}

\newtheorem{Thm}{Theorem}

\newtheorem{Cor}{Corollary}
\newtheorem{Prop}{Proposition}
\newtheorem{Def}{Definition}

\title{Born Geometry in a Nutshell}

\ShortTitle{Born Geometry in a Nutshell}

\author{Felix J. Rudolph$^*$\\
        Max-Planck-Institut f\"ur Physik, F\"ohringer Ring 6, 80805 Munich, Germany\\
        E-mail: \email{frudolph@mpp.mpg.de}}

\author{David Svoboda\footnote{Speaker}\\
        Perimeter Institute for Theoretical Physics, 31 Caroline St. N., Waterloo ON, N2L 2Y5, Canada\\
        E-mail: \email{dsvoboda@perimeterinstitute.ca}}

\abstract{We give a concise summary of the para-Hermitian geometry that describes a doubled target space fit for a covariant description of T-duality in string theory. This provides a generalized differentiable structure on the doubled space and leads to a kinematical setup which allows for the recovery of the physical spacetime. The picture can be enhanced to a Born geometry by including dynamical structures such as a generalized metric and fluxes which are related to the physical background fields in string theory.  We then discuss a generalization of the Levi-Civita connection in this setting -- the Born connection -- and twisting of the kinematical structure in the presence of fluxes.}

\FullConference{Corfu Summer Institute 2018 "School and Workshops on Elementary Particle Physics and Gravity"\\
		(CORFU2018)\\
		31 August - 28 September, 2018\\
		Corfu, Greece}

\begin{document}

\section{Introduction}
The interplay between physical concepts and geometric structures is at the heart of fundamental theoretical physics. The prime example is the role of Riemannian geometry in the formulation and understanding of general relativity. It is therefore natural to ask what novel geometric structures will be needed to encompass the ideas behind a theory of quantum gravity.

A promising candidate for such a theory is string theory. Due to the extended nature of the fundamental string and the presence of dualities, it has proven useful to work with a \emph{doubled geometry}. This is formed by extending spacetime to include extra coordinates corresponding to the winding modes of the string (whereas usual coordinates correspond to the momentum modes). This doubled geometry can be viewed as the target space of the fundamental string which is manifestly invariant under T-duality transformations exchanging winding and momentum modes. An effective action for this setup is provided by double field theory (DFT) \cite{Siegel:1993xq,Siegel:1993th,Hull:2009mi}. The doubled space can also be interpreted as the phase space of the fundamental string which then leads to the formulation of metastring theory \cite{Freidel:2014qna,Freidel:2015pka}.

The geometric structure of this doubled space is that of a \emph{para-Hermitian manifold} as was first demonstrated by Vaisman \cite{Vaisman:2012ke,Vaisman:2012px}. Such geometry provides the {\it kinematical structure} for the theory in the sense that it augments the doubled space with a differentiable structure, naturally adapted to the T-duality covariant setting. The dynamical data is then given by a generalized metric and generalized fluxes on the doubled geometry. Such a setup has been dubbed \emph{Born geometry} \cite{Freidel:2013zga}.

In this contribution we give an executive summary of para-Hermitian geometry in the context of the doubled geometry of string theory and how it is related to generalized geometry based on \cite{Freidel:2017yuv,Svoboda:2018rci,Freidel:2018tkj} (see \cite{Chatzistavrakidis:2018ztm} for related developments). We give the basic ingredients of Born geometry and present the unique torsion-free compatible connection for Born geometry. This can be seen as the analogy to the Levi-Civita connection for Riemannian geometry. We will also explain how the fluxes naturally arise in this setting. \textcolor{white}{\speaker{F.~J.~Rudolph, D.~Svoboda} }

\section{Generalized Kinematics}
It has been observed that the vector fields on the extended spacetime obey a new algebra relation which is different than the usual Lie algebra of vector fields given by the Lie bracket. One is therefore led to replace the Lie bracket with a new bracket operation called the {\it D-bracket} which leads to a redefinition of the notion of {\it differentiable structure} on the manifold, i.e. the Cartan calculus. All usual objects whose definitions use the Lie bracket, such as the torsion and Nijenhuis tensors, need to be replaced by their counterparts using the D-bracket. We call this new differentiable structure on the manifold the {\it generalized kinematical structure}. In this section we describe that such structure on an extended spacetime is naturally given by the data of a para-Hermitian manifold.

\subsection*{Para-Hermitian Geometry}
We now briefly introduce all important concepts of para-Hermitian geometry. For more details consult \cite{Cruceanu} and references therein.

\begin{Def}
Let $\PS$ be a $2n$-dimensional manifold. An almost para-Hermitian structure on $\PS$ is a pair $(\eta,K)$, where $\eta$ is a metric with signature $(n,n)$ and $K \in \text{End}(T\PS)$ with $K^2=\id$. These two are compatible in the sense that $K$ is an anti-isometry of $\eta$: $\eta(K\cdot,K\cdot)=-\eta(\cdot,\cdot)$.
\end{Def}

A consequence of the definition is that $K$ has two eigenbundles of rank $n$ corresponding to the eigenvalues $\pm 1$, which we will in the following denote by $L$ and $\Lt$. The fact that $L$ and $\Lt$ have the same rank makes $K$ an almost {\it para-complex} structure.  Another consequence of the definition is that the tensor $\omega \coloneqq \eta\circ K$ is a non-degenerate two-form, i.e. an {\it almost symplectic structure}, sometimes called the fundamental form. If $\rd \omega=0$, we call $(\eta,K)$ an (almost) {\it para-K\"ahler} structure.

It is useful to realize the analogy of para-Hermitian geometry with its complex counterpart, Hermitian geometry. Indeed, the only difference here is that the para-complex structure $K$ squares to $+\id$ as opposed to $-\id$ in the complex case. The isometry condition of a complex structure then also comes with the opposite sign. 

Just as in the complex case, we can invoke the notion of {\it integrability} which is governed by the {\it Nijenhuis tensor}:
\begin{align}\label{eq:nijenhuis}
\begin{aligned}
N_K(X,Y)&\coloneqq\frac{1}{4}\Big( [X,Y]+[KX,KY]-K([KX,Y]+[X,KY])\Big)\\
&=P[\Pt X,\Pt Y]+\Pt[P X,P Y],
\end{aligned}
\end{align}
where we introduced the projections onto $L$ and $\Lt$:
\begin{align}\label{eq:K-projections}
P=\frac{1}{2}(\id+K) \quad \mathrm{and} \quad \Pt=\frac{1}{2}(\id-K).
\end{align}
From the last line of \eqref{eq:nijenhuis}, it is obvious that $N_K$ vanishes if and only if both bundles $L$ and $\Lt$ are Frobenius integrable, i.e. closed under the Lie bracket. When this is the case, one gets local charts $X^I=(x^i,\xt^i)$, $i=1\cdots n$, such that $L=\text{span}\{\p_i=\frac{\p}{\p x^i}\}$ and $\Lt=\text{span}\{\pt^i=\frac{\p}{\p \xt_i}\}$ and the manifold looks locally like a product manifold.

We can also notice one of the most striking features of para-complex geometry which sets it apart from complex geometry -- the integrability of the eigenbundles $L$ and $\Lt$ is independent on each other, meaning that $L$ can be integrable while $\Lt$ is not and vice versa. This gives rise to the notion of {\it half-integrability}. In the subsequent discussion we will not reference integrability when it is not relevant and omit the ``almost'' prefix, while explicitly calling the para-Hermitian structure (half-)integrable whenever relevant.

\subsection*{The D-bracket}
We now explain how the para-Hermitian data $(\eta,K)$ naturally defines a D-bracket on the underlying manifold. The central theorem for this discussion is the following:

\begin{Thm}
Let $(\PS,\eta,K)$ be an almost para-Hermitian manifold. There exists a unique bracket operation $\bl\ , \ \br$: $\Gamma(T\PS)\times \Gamma(T\PS) \rightarrow \Gamma(T\PS)$ satisfying the following properties:
\begin{enumerate}
\item $\bl X,fY\br = f\bl X,Y\br+X[f]Y,\quad$ \hfill Leibniz property
\item $X[\eta(Y,Z)] = \eta(\bl X,Y\br,Z)+\eta(Y,\bl X,Z\br)$ \hfill  Compatibility with $\eta$ \\
$\eta(Y,\bl X,X\br ) = \eta(\bl Y,X\br,X),\quad$ 
\item $\mathcal{N}_K(X,Y)=\bl X,Y\br +\bl K X,K Y\br  - K\big(\bl K X,Y\br  + \bl X,K Y\br\big)=0,\quad$ \hfill Compatibility with $K$
\item $
\bl PX,PY \br = P([PX,PY]),\quad\bl \Pt X,\Pt Y \br =\Pt([ \Pt X,\Pt Y]),$ \hfill Relationship with the Lie bracket.
\end{enumerate}
for any vector fields $X,Y\in\Gamma(T\PS)$ and any function $f\in C^\infty(\PS)$. 
\end{Thm}
In the above statement we introduced the {\it generalized Nijenhuis tensor} of $K$, $\mathcal{N}_K$. One can glean that this tensorial quantity is nothing else than the usual Nijenhuis tensor, except the Lie bracket is replaced with the D-bracket.

The following statement shows that the D-bracket not only exists, but can be easily written out in a convenient form:

\begin{Prop}
The D-bracket on a para-Hermitian manifold $(\PS,\eta,K)$ is given by 
\begin{align}\label{eq:D-bracket_canonical}
\eta(\bl X,Y\br,Z)=\eta(\nc_XY-\nc_YX,Z)+\eta(\nc_ZX,Y),
\end{align}
where $\nc$ is the {\it canonical connection}, which can be defined via the Levi-Civita connection $\lc$ of $\eta$:
\begin{align*}
\nc_XY=P\lc_X(PY)+\Pt\lc_X(\Pt Y).
\end{align*}
\end{Prop}
In the expression \eqref{eq:D-bracket_canonical} we can notice that the first two terms, which are skew-symmetric under the exchange of $X$ and $Y$, are reminiscent of how one can define a Lie bracket using a torsionless connection. Even though $\nc$ is not torsionless, in the limit when $(\eta,K)$ is para-K\"ahler, it is easy to check that $\nc=\lc$ and therefore $\nc_XY-\nc_YX=[X,Y]$.

We should also point out that the D-bracket in particular coincides with the D-bracket appearing in the physics literature, where this operation is considered usually on flat manifolds, $\mathbb{R}^{2n}$ or tori $T^{2n}$, both of which are simple examples of flat para-Hermitian manifolds.
\begin{Cor}
On a flat para-Hermitian manifold, the D-bracket takes the form 
\begin{align*}
\bl X,Y \br=\left(X^I\p_IY^J-Y^I\p_IX^J+\eta_{IL}\eta^{KJ}Y^I\p_KX^L\right)\p_J,
\end{align*}
where $\p_I=(\p_i,\pt^i)$ are the partial derivatives along the coordinates $(x^i,\tl{x}_i)$.
\end{Cor}

We emphasize here that even though the D-bracket is unique, the choice of connection to express it via the formula \eqref{eq:D-bracket_canonical} is not unique and $\n^c$ is only one choice of such connection. In \cite{Svoboda:2018rci}, the set of connections that yield the D-bracket are called {\it adapted}.

\subsection*{Generalized Torsion}
The usual torsion of a linear connection $\n$ can be understood as a tensorial quantity measuring how much the skew combination $\n_XY-\n_YX$ deviates from the Lie bracket, $[X,Y]$. The same is true for a so-called generalized torsion. In the light of the equation \eqref{eq:D-bracket_canonical}, we define

\begin{Def}
Let $(\PS,\eta,K)$ be a para-Hermitian manifold and $\n$ a linear connection. The {\it generalized torsion} of $\n$ is the tensorial quantity defined by
\begin{align*}
\mathcal{T}^\n(X,Y,Z)\coloneqq \eta(\n_XY-\n_YX,Z)+\eta(\n_ZX,Y)-\eta(\bl X,Y\br,Z)
\end{align*}
\end{Def}

From this definition we see that it precisely captures how much does the cyclic combination of $\n$, $\eta(\n_XY-\n_YX,Z)+\eta(\n_ZX,Y)$ deviate from the D-bracket. If $\mathcal{T}^\n=0$, $\n$ via this formula defines  a D-bracket.

\subsection*{Courant Algebroids and the Relationship to the D-bracket}
In this section we explore how the D-bracket relates to the Dorfman bracket. We explain that when $K$ is integrable, the D-bracket can be seen as a sum of two Dorfman brackets corresponding to certain Courant algebroids. In particular, this equips the tangent bundle of the para-Hermitian manifold with two natural Courant algebroid structures defined purely in terms of the para-Hermitian data.

We start off with a definition of a Courant algebroid.
\begin{Def}
Let $E\rightarrow M$ be a vector bundle. A Courant algebroid is a quadruple $(E,a,\langle\ , \ \rangle_E,\bl\ ,\ \br)$, where $a$ is bundle map $E\rightarrow TM$ called the anchor, $\langle\ , \ \rangle_E:\se(E)\times\se(E)\rightarrow C^\infty(M)$ is a non-degenerate symmetric pairing and $\bracd:\se(E)\times\se(E)\rightarrow \se(E)$ is a bracket operation called the Dorfman bracket\footnote{The Dorfman bracket is sometimes called the Dorfman derivative or generalized Lie derivative.}, such that
\begin{enumerate}
\item $a(X)\langle Y,Z\rangle_E=\la\bl X,Y\br,Z\ra_E+\la Y,\bl X,Z\br\ra_E$
\item $\langle\bl X,X\br,Y\rangle_E=\frac{1}{2}a(Y)\langle X,X\rangle_E$
\item $\bl X, \bl Y,Z\br\br=\bl\bl X,Y\br, Z\br+\bl Y,\bl X,Z\br\br$,
\end{enumerate}
\end{Def}
A canonical example is given by the {\it standard Courant algebroid}, where $E=(\TT)M$, $a:(\TT)M\rightarrow TM$ is the natural projection, the pairing is given by $\langle x+\ap ,y+\bt \rangle=\ap(y)+\bt(x)$ and the bracket operation is the Dorfman bracket:
\begin{align*}
\bl x+\ap,y+\bt \br = [x,y]+\Lie_x\bt+\imath_y\rd \ap.
\end{align*}
where $x,y\in\Gamma(TM)$, $\ap,\bt\in\Gamma(T^*M)$ and $\Lie_x$ is the Lie derivative. We further introduce the following notation:
\begin{Def}
Let $\bracd$ be the D-bracket on a para-Hermitian manifold. Then the {\it projected brackets} $\bracd_P$ and $\bracd_\Pt$ are defined as follows
\begin{align*}
\eta(\bl X,Y\br_P,Z)&=\eta(\nc_{PX}Y-\nc_{PY}X,Z)+\eta(\nc_{PZ}X,Y)\\
\eta(\bl X,Y\br_\Pt,Z)&=\eta(\nc_{\Pt X}Y-\nc_{\Pt Y}X,Z)+\eta(\nc_{\Pt Z}X,Y).
\end{align*}
The brackets are related by $\bl X,Y\br=\bl X,Y\br_P+\bl X,Y\br_\Pt$.
\end{Def}

We continue with the following simple observation. Since $L$ and $\Lt$ are isotropic with respect to $\eta$, whenever $\xt \in \se(\Lt)$, $\eta(\xt,\cdot)$ is an element of $\se(L^*)$ because it contracts only with vectors in $L$. In fact, any section of $L^*$ is of this form by non-degeneracy of $\eta$. We therefore have the following vector bundle isomorphism:
\begin{align*}
\begin{aligned}
\rho:\ T\PS &= L\oplus \Lt\rightarrow L\oplus L^*\\
X &= x+\xt \mapsto x+\eta(\xt).
\end{aligned}
\end{align*}
and similarly we can define $\tl{\rho}:L\oplus \Lt \rightarrow \Lt \oplus \Lt^*$. Moreover, when $L$ is an integrable distribution (i.e. whenever $\Pt[PX,PY]=0\ \forall X,Y \in \Gamma(T\PS)$), the manifold $\PS$ is {\it foliated} by half-dimensional manifolds $\ell_i$ called leaves. This means that there is a partition of the $2n$-dimensional manifold $\PS$ into $n$-dimensional manifolds as $\PS=\bigcup_i \ell_i$. We can therefore view $\PS$ as an $n$-dimensional manifold (given by the union of the leaves $\ell_i$ of the foliation) and to avoid confusion we denote this $n$-dimensional manifold as $\mathcal{F}$. Then, by definition, $L=T\mathcal{F}$ and $T\PS=L\oplus L^*=(\TT)\mathcal{F}$ in a natural way. It turns out that the projected bracket $\bracd_P$ is nothing else than the Dorfman bracket of the standard Courant algebroid $(\TT)\mathcal{F}$ mapped under the map $\rho$:

\begin{Thm}
Let $(\PS,\eta,K)$ be an almost para-Hermitian manifold and $\bracd$, $\bracd_P,\bracd_\Pt$ the associated D-bracket and projected brackets, respectively. Whenever $L$ is integrable, $T\PS$ acquires the structure of a Courant algebroid $(T\PS,P,\eta,\bracd_P)$. The map $\rho$ is then an isomorphism of Courant algebroids:
\begin{align*}
(T\PS,P,\eta,\bracd_P) \overset{\rho_+}{\longrightarrow}((\TT)\mathcal{F},a,\la\ ,\ \ra,\bracd_{\mathcal{F}}),
\end{align*}
where $((\TT)\mathcal{F},a,\la\ ,\ \ra,\bracd_{\mathcal{F}})$ is the standard Courant algebroid of $\mathcal{F}$. This means that
\begin{align*}
\rho \bl X,Y\br_P=\bl \rho X,\rho Y\br_\mathcal{F}, \quad \eta(X,Y)=\la \rho X,\rho Y\ra, \quad \rho\circ P=a\circ \rho.
\end{align*}
An analogous statement holds for $\Lt$ where the Courant algebroid structure is then given by \\ $(T\PS, \Pt, \eta, \bracd_\Pt)$.
\end{Thm}

\section{Born Geometry}
\label{sec:borngeo}

We have seen that para-Hermitian geometry is a para-complex geometry $(\PS,K)$ with a compatible neutral metric $\eta$, i.e. one of signature $(n,n)$. We have also seen how if $L$ is integrable we can recover an $n$-dimensional manifold $\mathcal{F}$ with $L=T\mathcal{F}$ and and a standard Courant algebroid structure which can be identified with the physical spacetime. In order to make connection with a string background, we also need to recover the physical background fields on $\mathcal{F}$, i.e. the spacetime metric $g$ and the Kalb-Ramond field $B$.

Therefore the next step is to add another compatible metric $\HH$ of signature $(2n,0)$ to the picture, giving rise to what has been named Born geometry \cite{Freidel:2013zga}. The additional Riemannian structure $\HH$ defines two more endomorphisms of the tangent bundle which -- along with the para-complex structure $K$ already in place -- form a para-quaternionic structure \cite{Freidel:2015pka,Freidel:2013zga,Ivanov:2003ze}.

\begin{Def}
Let $(\PS,\eta,\omega)$ be a para-Hermitian manifold and let $\HH$ be a Riemannian metric satisfying
\begin{align}
\eta^{-1}\HH=\HH^{-1}\eta,\quad \omega^{-1}\HH=-\HH^{-1}\omega .
\end{align}
Then we call the triple $(\eta,\omega,\HH)$ a Born structure on $\PS$ where $\PS$ is called a Born manifold and $(\PS,\eta,\omega,\HH)$ a Born geometry.
\end{Def}
Note that in this definition we view $(\eta,\omega,\HH)$ as maps $T\PS\to T^*\PS$. One can show \cite{Freidel:2018tkj} that there always exists a choice of frame on $T\PS=L\oplus\Lt$ where the generalized metric takes the form 
\begin{equation}
\HH = \begin{pmatrix}
g & 0 \\ 0 & g^{-1}
\end{pmatrix}
\label{eq:genmetric}
\end{equation}
where $g$ is a Riemannian metric on $L=T\mathcal{F}$. The appearance of the B-field will be discussed below in the context of twisting the para-Hermitian structure. 

In order to understand the nature of a Born structure we review its three fundamental structures. First, as we have seen, it contains an almost {\it para-Hermitian} structure $(\omega, K)$ with compatibility
\be
K^2=\id,\qquad \omega(KX,KY)=-\omega(X,Y). 
\ee
Next, the compatibility between $\eta$ and $\HH$ implies that $J=\eta^{-1}\HH \in \mathrm{End}\, T\PS$ defines what we refer to as a  {\it  chiral} structure\footnote{The projectors $P_\pm=\frac12(\id\pm J)$ define, similarly to projectors $P$, $\Pt$, a splitting of the tangent bundle $T\PS = C_+ \oplus C_-$ which mirrors the right/left chiral splitting of string theory. In other words,
given a worldsheet $X:\Sigma \to \PS$ with $\Sigma$ a Riemann surface, we have that $\pa_z X$ takes values in $C_+$ and $\pa_{\bar{z}} X$ in $C_-$. Hence $J$ is called a chiral structure.} $(\eta,J)$ on $\PS$
\begin{align}
J^2=\id,\qquad 
\eta(JX,JY)=\eta(X,Y).
\end{align}
Finally, the compatibility between  $\HH$ and $\omega$ defines an almost {\it Hermitian} structure $(\HH,I)$ on $\PS$
\be
I^2=-\id,\qquad \HH(IX,IY)= \HH(X,Y).
\ee
These three structures, para-Hermitian, chiral and Hermitian satisfy in turn an extra compatibility equation
\be
KJI=\id,
\ee
which follows directly from their definition $K=\omega^{-1}\eta$, $J=\eta^{-1} \HH$, $I=\HH^{-1}\omega$. This means that the triple $(I,J,K)$  form an almost para-quaternionic structure 
\begin{align}
-I^2=J^2=K^2=\id, \qquad 
\{I,J\}=\{J,K\}=\{K,I\}=0,\qquad 
KJI=\id,
\end{align}
where $\{\ , \ \}$ is the anti-commutator. 

\begin{table}[h!]
\renewcommand{\arraystretch}{1.3}
\centering
\resizebox{0.9\textwidth}{!}{%
\begin{tabular}{@{}ccc@{}}
\toprule
$I={\HH}^{-1}{\omega}=-\omega^{-1}\HH$ & $J={\eta}^{-1}{\HH}={\HH}^{-1}{\eta}$ & $K={\eta}^{-1}{\omega}={\omega}^{-1}{\eta}$  \vspace{8pt} \\
$-I^2=J^2=K^2=\id$                     & $\{I,J\}=\{J,K\}=\{K,I\}=0$           & $IJK=-\id$     \vspace{8pt} \\
$\HH(IX,IY)=\HH(X,Y)$              & $\eta(IX,IY)= -\eta(X,Y)$         & $\omega(IX,IY)=\omega(X,Y)$             \\
$\HH(JX,JY)=\HH(X,Y)$              & $\eta(JX,JY)= \eta(X,Y)$          & $\omega(JX,JY)=-\omega(X,Y)$            \\
$\HH(KX,KY)=\HH(X,Y)$              & $\eta(KX,KY)= -\eta(X,Y)$         & $\omega(KX,KY)=-\omega(X,Y)$            \\ \bottomrule
\end{tabular}%
}
\caption{Summary of structures in Born geometry. Here $\{\ , \ \}$ is the anti-commutator.}
\label{tab:BornGeo}
\end{table}

\section{Born Connection}
We have now all ingredients in place to examine the Born connection. It is the unique connection compatible with the Born geometry, i.e. with the three defining structures $(\eta,\omega,\HH)$, which has no generalized torsion. The Born connection can be seen as the analogue of the Levi-Civita connection which is the unique, torsionfree, metric-compatible connection for Riemannian geometry.

\begin{Thm}\label{th:BornConnection}
Let $(\PS,\eta,\omega,\Hh)$ be a Born geometry with $K=\eta^{-1}\omega$ the corresponding almost para-Hermitian structure. Then there exists a unique connection called the Born connection denoted by $\nB$ which
\begin{itemize}
  \item is compatible with the Born geometry $(\PS,\eta,\omega,\Hh)$,
  \item has a vanishing generalized torsion $\Th=0$.
\end{itemize}
This connection can be explicitly expressed in terms of the three defining structures $(\eta,\omega,\Hh)$ of the Born geometry and the canonical D-bracket. It can be concisely written in terms of the para-Hermitian structure $K$ and the chiral projections $X_\pm=\frac{1}{2}(\id\pm J)X$ as
\begin{equation}
\nB_XY = \bl X_-,Y_+ \br_+ + \bl X_+,Y_-\br_- + (K\bl X_+,KY_+ \br)_+ + (K\bl X_-,KY_-\br)_- .
\label{eq:BornBracket}
\end{equation}
\end{Thm}

The theorem is proven in \cite{Freidel:2018tkj}.

\section{Fluxes}
We have seen in previous sections that any para-Hermitian structure $(\eta, K)$ gives rise to its unique associated D-bracket. In this section we will explain how the {\it twisted} D-bracket appears in this context. We will see that the fluxes twisting the bracket appear when we choose a different para-Hermitian structure $K'$ compatible with the same $\eta$, and express its associated D-bracket in the splitting given by the original $K$. The appearance of fluxes can be therefore intuitively understood as a choice of a splitting not compatible with the D-bracket at hand and the fluxes can be seen as certain obstructions of an integrability of this splitting.

We now define a transformation of the para-Hermitian structure analogous to a B-field transformation in generalized geometry.

\begin{Def}
Let $(\eta,K)$ be a para-Hermitian structure on $\PS$. A B-transformation of $K$ is given by
\begin{align*}
K\overset{e^{B}}{\longmapsto} K_{B}=e^{B} K e^{-{B}},\quad
e^{B}\coloneqq
\begin{pmatrix}
\id & 0 \\
B & \id
\end{pmatrix} \in \text{End}(T\PS)
\end{align*}
where the map $e^B$ is expressed in the matrix representation corresponding to the splitting $L\oplus L$ and $B:L\rightarrow \Lt$ is a skew map such that $\eta(BX,Y)=-\eta(X,BY)$.
\end{Def}
It is easy to see that ${B}$ can be given by either a two-form $b$ or a bivector $\beta$,
\begin{align}\label{eq:b-beta}
\eta(BX,Y)=b(X,Y)=\beta(\eta (X),\eta (Y)),
\end{align}
where $b$ is of type $(+2,-0)$ and $\beta$ is of type $(+0,-2)$, so we can write $b(X,Y)=b(\tl{x},\tl{y})$. In coordinates, we have
\begin{align*}
b=b_{ij}\dd x^i\wedge \dd x^j,\quad \beta=\beta_{ij} \pt^i\wedge \pt^j.
\end{align*}

It can be checked that $K_B$, the B-transformation of $K$, is a new (almost) para-Hermitian structure. $K$ and $K_B$ share the same $-1$ eigenbundle, but the $+1$ eigenbundle is sheared by the map $B$: $L\mapsto L_B=L+B(L)$. The B-transformation therefore in general does not preserve integrability of the para-Hermitian structure.

When the starting para-Hermitian structure is in fact an integrable para-K\"ahler structure, the D-bracket associated to $K_B$ carries with respect to $K$ well-known fluxes from physical literature:

\begin{Prop}
Let $K_B$ be a B-transformation of a para-K\"ahler structure $(\PS,\eta,K)$. Then the D-bracket associated to $K_B$ is given by
\begin{align}\label{twistedDbrac}
\eta(\bl X,Y\br^{B},Z)=\eta(\bl X,Y\br^D ,Z)-(\rd b)(X,Y,Z).
\end{align}
where $\bl\ ,\ \br$ denotes the D-bracket of $K$. The different components of $\rd b$ with respect to the splitting of $K_B$ yield 
\begin{align*}
\rd b^{(3,0)_B}&=\hat{H}=H+\tl{R}\\
\rd b^{(3,0)_B}&=\tl{Q},
\end{align*}
where $\hat{H}$ is the Covariantized H-flux, $H$ is the usual H-flux and $\tl{R}$ and $\tl{Q}$ are the dual R- and Q-fluxes.
\end{Prop}

We see the appearance of the {\it dual} fluxes, meaning $\tl{R}$ is a three-vector on $\tl{L}$, while the usual R-flux is usually understood as a three-vector on $L$, i.e. $R$ and $\tl{R}$ have an opposite index structure and similarly with $Q$ and $\tl{Q}$. If we wanted to achieve the usual $Q$ and $R$, we would have to perform the dual B-transformation, shearing $\Lt\mapsto \Lt+B(\Lt)$. Indeed, this has been done in \cite{Marotta:2018myj}, where it was additionally also shown that the simultaneous B-transformation shearing both in $L$ and $\Lt$ directions yields the whole hierarchy of fluxes appearing in DFT. 

Let us conclude this section by giving the effect of a B-transformation on the generalized metric $\HH$ as given in \eqref{eq:genmetric}. The transformed generalized metric takes the form
\begin{equation}
\HH \overset{e^{B}}{\longmapsto} e^{-B}\HH e^{-B} = 
\begin{pmatrix}
g - b g^{-1} b& bg^{-1} \\ -g^{-1}b & g^{-1}
\end{pmatrix}
\end{equation}
which is familiar from generalized geometry and DFT.

\section{Conclusion}
The construction of a para-Hermitian manifold with a corresponding foliation (when one of the subdistributions is integrable) to recover the physical spacetime of a string theory background has been summarized here. This puts the idea of a doubled target space for the fundamental string on a firm mathematical base and spells out the precise relation to generalized geometry and double field theory. 

On top of this generalized kinematical structure one can include the dynamical fields via a generalized metric which corresponds to a choice of Riemannian metric on the subspace, giving rise to a Born geometry. The generalized fluxes appear via a twisting of the para-Hermitian structure and hence the D-bracket. This allows for an inclusion of the B-field and subsequently the other fluxes of string theory.

Concrete examples of para-Hermitian and Born geometries such as $\PS = TM$, i.e. where the doubled space is the tangent bundle of some manifold, can be found in \cite{Marotta:2018myj}. Group manifolds such as Drinfel'd doubles can also be viewed from a para-Hermitian perspective and provide another class of examples \cite{Marotta:2018myj,Mori:2019slw,inprep}.

In future work, we aim to develop this formalism further and show how it applies in various well-known settings where T-duality arises, such as topological T-duality \cite{tduality1,tduality2,tduality3}.

\section*{Acknowledgments}
The authors would like to thank the organizers of the ``Corfu Summer Institute 2018'' for the invitation. F.J.R. is supported by the Max-Planck-Society. The work of D.S. is supported by NSERC Discovery Grant 378721.

\end{document}